\documentclass[11pt]{article}
\usepackage{amssymb}\usepackage{amsmath}
\usepackage[all]{xy}
\usepackage[dvips]{graphicx}

\numberwithin{equation}{section}

\def\be{\begin{equation}}
\def\ee{\end{equation}}
\def\ba{\begin{align}}
\def\ea{\end{align}}

\def\yboxit#1#2{\vbox{\hrule height #1 \hbox{\vrule width #1
\vbox{#2}\vrule width #1 }\hrule height #1 }}
\def\fillbox#1{\hbox to #1{\vbox to #1{\vfil}\hfil}}
\def\ybox{{\lower 1.3pt \yboxit{0.4pt}{\fillbox{8pt}}\hskip-0.2pt}}
\def\comments#1{}

\def\half{\frac{1}{ 2}}

\def\ket#1{|#1\rangle}

\def\cK{{\cal K}}

\def\CN{{\cal N}}

\def\CC{{\cal C}}

\def\CC{{\cal C}}

\def\II{\relax{I\kern-.10em I}}

\def\IZ{\relax\ifmmode\mathchoice
{\hbox{\cmss Z\kern-.4em Z}}{\hbox{\cmss Z\kern-.4em Z}}
{\lower.9pt\hbox{\cmsss Z\kern-.4em Z}}
{\lower1.2pt\hbox{\cmsss Z\kern-.4em Z}}\else{\cmss Z\kern-.4em
Z}\fi}
\def\IB{\relax{\rm I\kern-.18em B}}
\def\IC{{\relax\hbox{$\inbar\kern-.3em{\rm C}$}}}
\def\ID{\relax{\rm I\kern-.18em D}}
\def\IE{\relax{\rm I\kern-.18em E}}
\def\IF{\relax{\rm I\kern-.18em F}}
\def\IG{\relax\hbox{$\inbar\kern-.3em{\rm G}$}}
\def\IGa{\relax\hbox{${\rm I}\kern-.18em\Gamma$}}
\def\IH{\relax{\rm I\kern-.18em H}}
\def\II{\relax{\rm I\kern-.18em I}}
\def\IK{\relax{\rm I\kern-.18em K}}
\def\IN{\relax{\rm I\kern-.18em N}}
\def\IP{\relax{\rm I\kern-.18em P}}

\def\inbar{\,\vrule height1.5ex width.4pt depth0pt}

\font\cmss=cmss10 \font\cmsss=cmss10 at 7pt
\def\IR{\relax{\rm I\kern-.18em R}}

\def\ch{{\rm ch}}

\def\CN{{\cal N}}

\def\lp10{l_P^{10}}
\def\lp11{l_P^{11}}
\def\R11{R_{11}}

\def\ch{{\rm ch}}

\font\manual=manfnt
\def\dbend{\lower3.5pt\hbox{\manual\char127}}

\def\IZ{\relax\ifmmode\mathchoice
{\hbox{\cmss Z\kern-.4em Z}}{\hbox{\cmss Z\kern-.4em Z}}
{\lower.9pt\hbox{\cmsss Z\kern-.4em Z}} {\lower1.2pt\hbox{\cmsss
Z\kern-.4em Z}}\else{\cmss Z\kern-.4em Z}\fi}
\def\half {\frac{1}{ 2}}

\def\bar{\overline}

\def\CN{{\cal N}}

\def\rt2{\sqrt{2}}
\def\irt2{\frac{1}{\sqrt{2}}}

\font\cmss=cmss10
\font\cmsss=cmss10 at 7pt
\def\IL{\relax{\rm I\kern-.18em L}}
\def\IH{\relax{\rm I\kern-.18em H}}
\def\IR{\relax{\rm I\kern-.18em R}}
\def\inbar{\vrule height1.5ex width.4pt depth0pt}
\def\IC{\relax\hbox{$\inbar\kern-.3em{\rm C}$}}
\def\rlx{\relax\leavevmode}
\def\ZZ{\rlx\leavevmode\ifmmode\mathchoice{\hbox{\cmss Z\kern-.4em Z}}
 {\hbox{\cmss Z\kern-.4em Z}}{\lower.9pt\hbox{\cmsss Z\kern-.36em Z}}
 {\lower1.2pt\hbox{\cmsss Z\kern-.36em Z}}\else{\cmss Z\kern-.4em
 Z}\fi}
\def\IZ{\relax\ifmmode\mathchoice
{\hbox{\cmss Z\kern-.4em Z}}{\hbox{\cmss Z\kern-.4em Z}}
{\lower.9pt\hbox{\cmsss Z\kern-.4em Z}}
{\lower1.2pt\hbox{\cmsss Z\kern-.4em Z}}\else{\cmss Z\kern-.4em
Z}\fi}

\def\ch{{\rm ch}}

\font\manual=manfnt
\def\dbend{\lower3.5pt\hbox{\manual\char127}}

\def\IZ{\relax\ifmmode\mathchoice
{\hbox{\cmss Z\kern-.4em Z}}{\hbox{\cmss Z\kern-.4em Z}}
{\lower.9pt\hbox{\cmsss Z\kern-.4em Z}} {\lower1.2pt\hbox{\cmsss
Z\kern-.4em Z}}\else{\cmss Z\kern-.4em Z}\fi}
\def\half {\frac{1}{ 2}}

\def\bar{\overline}

\def\rt2{\sqrt{2}}
\def\irt2{\frac{1}{\sqrt{2}}}

\def\T{\widetilde}

\def\sNSNS{\text{\tiny{NSNS}}}
\def\sRR{\text{\tiny{RR}}}
\def\sNS{\text{\tiny{NS}}}
\def\sR{\text{\tiny{R}}}
\def\sId{\text{\tiny{Id}}}
\def\scont{\text{\tiny{cont}}}

\title{\Large{\bf D-branes in Unoriented Non-critical Strings and \\
Duality in $SO(N)$ and $Sp(N)$ Gauge Theories}}
\author{Sujay K. Ashok$^{a}$, Sameer Murthy$^{b}$ and Jan Troost$^{c}$ }
\date{}
\begin{document}
\maketitle

\begin{center}
$^{a}$Perimeter Institute for Theoretical Physics\\
Waterloo, Ontario, ON N$2$L$2$Y$5$, Canada \\
$^{b}$Abdus Salam International Center for Theoretical Physics \\
Strada Costiera 11, Trieste, $34014$, Italy \\
$^{c}$Laboratoire de Physique Th\'eorique\footnote{Unit\'e Mixte du CRNS et de l'Ecole
  Normale
Sup\'erieure associ\'ee \`a l'universit\'e Pierre et Marie Curie 6, UMR
8549. Preprint LPTENS-07/10.}, Ecole Normale Sup\'erieure \\
$24$ Rue Lhomond Paris $75005$, France
\end{center}
\begin{abstract}
We exhibit exact conformal field theory descriptions of $SO(N)$ and $Sp(N)$ pairs of Seiberg-dual gauge theories within string theory. The ${\cal N}=1$ gauge theories with flavour are realized as low energy limits of the worldvolume theories on D-branes in unoriented non-critical superstring backgrounds. These unoriented backgrounds are obtained by constructing exact crosscap states in the $SL(2,\mathbb{R})/U(1)$ coset conformal field theory using the modular bootstrap method. Seiberg duality is understood by studying the behaviour of the boundary and crosscap states under monodromy in the closed string parameter space.
\end{abstract}

\section{Introduction}
Electric-magnetic duality of ${\cal N}=1$ gauge theories in four dimensions is
an important tool to understand supersymmetric generalizations of quantum
chromodynamics \cite{Seiberg:1994pq}. For instance, it has recently been used
to argue for the {\em generic} existence of long-lived meta-stable
supersymmetry breaking vacua in supersymmetric gauge theories
\cite{Intriligator:2006dd}. In order to further exploit the fruitful interplay
between gauge theory and string theory, it is natural to realize Seiberg
duality within a brane set-up in string theory \cite{HananyW, ElitzurGK}. To sharpen the embedding map, it is useful to incorporate the
backreaction of the heaviest, NS5-branes, used within the brane set-up into the
background geometry. It is possible to do this in a double scaling limit
\cite{GiveonKP, GiveonK}. The D-branes that realize the gauge theories can
then be described by an exact boundary conformal field theory, and one can
analyze Seiberg duality exactly within string theory, as argued and realized
in \cite{MurthyT} for supersymmetric QCD with $SU(N_c)$ gauge group. It is
worth trying to extend this to $SO$ and $Sp$ gauge theories, since both the bulk
and boundary conformal field theories, as well as the
gauge theories, realize new physical phenomena.

In the rest of the introduction we review Seiberg duality in $SO/Sp$ gauge theories \cite{IntriligatorS} in the context of NS$5$/D-brane set-ups in type IIA string theory, in as far as it is useful for our present purposes. We then move to a worldsheet description of the backreacted theory in a double scaling limit in section \ref{worldsheet}. The main  technical result in this section is the computation of the supersymmetric crosscap states built on the representations of the $\CN=2$ superconformal algebra.
In the modular bootstrap procedure used to construct D-branes in Liouville
theory \cite{FateevZZ, Zamolodchikov:2001ah} and their supersymmetric
extensions, \cite{EguchiS, AhnRSone, AhnRStwo, janbranes}, the crosscap states are found by
requiring that, in the overlap of the crosscaps with the localized 
brane, the $\Omega$-twisted characters of the superconformal algebra are obtained in the open string channel. We give a closed string orientifold description of the different crosscaps by computing the Klein-bottle amplitudes. Some technical details regarding the amplitudes are collected in the appendices. Crosscap states in the cigar background have already been studied
in detail in a ``continuum limit" \cite{Nakayama}. However, for our purposes, it is necessary to require integral $U(1)_R$ charges so as to be able to GSO project.

In section \ref{flip}, we use these crosscaps along with the boundary states
described in \cite{FotopoulosNP, AshokMT} to engineer a $SO/Sp$ gauge theory
in four dimensions with flavours in the fundamental representation. We study
how electric-magnetic or Seiberg duality is realized in the gauge theory from
a worldsheet perspective, following the ideas in \cite{MurthyT}. This involves
understanding how boundary states and crosscaps behave under a monodromy in
the closed string parameter space. We find that within the context of a smooth superstring 
background,  our results realize 
the results obtained in the geometric type IIA set-up involving D$4$ branes
and NS$5$ branes \cite{ElitzurGK} and are consistent with field theory
expectations. We conclude with a brief discussion in section
\ref{conclusions}.

\subsection{Seiberg Duality from NS$5$/D-Brane Set-ups}\label{geometric}

When tackling the exact description of Seiberg duality in the double scaling limit, it will be useful to keep in mind the bulk configuration before taking the limit. We review briefly how Seiberg duality was analyzed by studying the low energy theory on D$4$ branes suspended between the NS$5$ branes in type IIA string theory. The realization of Seiberg duality in this brane set up has been 
discussed in detail in the literature \cite{ElitzurGK, ElitzurGKtwo, EvansJS}.
As we will show, some aspects of this description of Seiberg duality are 
closely related to our worldsheet discussion. 

\begin{figure}[h]
\includegraphics[scale=.7]{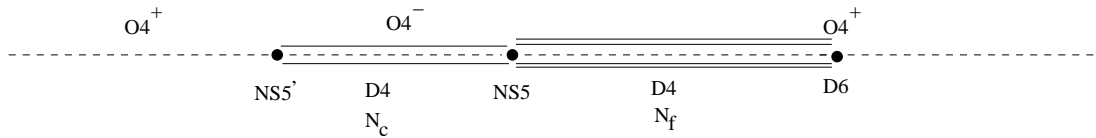} 
\caption{Electric brane set-up, including an orientifold four-plane that 
changes type, and $N_c$ colour D4-branes and $N_f$ flavour D4-branes ending on 
D6-branes. \label{orient4}}
\end{figure}

We begin with the electric set-up as shown in figure \ref{orient4}. We have shown the position of the various objects in the $x^6$
direction. The O$4$ plane extends all along the $x^6$ direction. Note that
when the O$4$ plane crosses an NS$5$ brane, it goes from the positively
charged $+$ type to the negatively charged $-$ type.

Flavour is realized by D$6$ branes to the right of the NS brane as shown in
figure \ref{orient4}. Since the low-energy gauge theory physics is independent
of the position of the D$6$ brane, one can also take the limit when it is at
infinity. This limit has an interesting relation to the geometric engineering
of gauge theories using D$6$ branes wrapped on the $3$-sphere
\cite{BershadskyVS, OV} in the deformed conifold. Seiberg duality in that
context was implemented by the so-called $\mu-$transition. The deformed
conifold is defined as the zero locus
\begin{equation}
x^2+y^2+z^2+u^2 = \mu \qquad \text{in}\quad \mathbb{C}^4 \,.
\end{equation}
The $\mu-$transition amounts to the operation $\mu\rightarrow -\mu$. In the
absence of flavour branes, such transitions have been studied in detail in
\cite{HoriH} by lifting the brane and orientifolds to M-theory following
earlier work in \cite{AtiyahW}.

Now the NS$5$ brane configuration we started with is related to the geometric
set-up by T-dualities \cite{OVKthree, GiveonK}. It is known that the closed
string parameter $\mu$ of the conifold maps to the relative position of the NS
and NS' branes. Electric-magnetic duality is therefore implemented in the IIA
brane set-up by exchanging the positions of the NS' and NS along the $x^6$
direction. The presence of the orientifold implies that we necessarily
pass through a strong coupling region in order to implement Seiberg duality
\cite{EvansJS}. 

\begin{figure}[h]
\includegraphics[scale=.7]{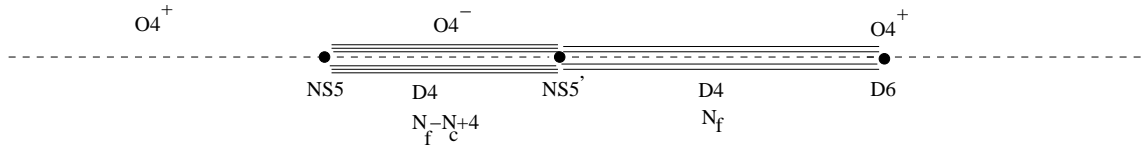} 
\caption{Magnetic brane set-up, including an orientifold four-plane that 
changes type, and $N_f-N_c+4$ colour D4-branes and $N_f$
flavour D4-branes ending on 
D6-branes. \label{orient4magn}}
\end{figure}

In order to derive the magnetic configuration, one uses the fact that a
certain linking number \cite{HananyW} must be conserved in the duality
transition. For the NS$5$ brane, the linking number is defined for each
NS$5$-brane as \cite{HananyW, ElitzurGK}
\begin{equation}\label{linking}
l_{NS}= \half(R_{D6}-L_{D6})+(L_{D4}-R_{D4})+Q(O4)(L_{O4}-R_{O4}) \,,
\end{equation}
where $R(X)$ ($L(X)$) refer to the number of branes of type $X$ to the right
(left) of the NS$5$ brane. For the electric configuration, which leads to a
$SO(N_c)$ theory with $N_f$ flavours, the linking number for the NS brane is
equal to
\begin{equation}
l_{NS}=-\half\, N_{f}+N_c-2\,.
\end{equation}
We count charges such that $N_c$ D$4$ branes lead to a colour group of
$SO(N_c)$. We have shown the magnetic configuration in 
figure \ref{orient4magn}.

Note that we are forced to include an additional $4$ D$4$ branes in
order to leave invariant the linking number. One can check that this brane
configuration has the correct properties of the Seiberg dual gauge theory,
including the presence of the $\half N_{f}(N_{f}+1)$ mesonic fields (the fluctuation of the $D4$ branes stretched between the NS' and D$6$) and the superpotential term $Mq\tilde{q}$.

It is the equality of the linking number in the electric and magnetic
configurations that forces us to include the $4$ extra D$4$ branes that
realize the colour gauge theory. It is not possible to derive the generation
of the $4$ extra D$4$ branes in this semi-classical brane set-up. See however
\cite{EvansJS} for a derivation of this phenomenon from a strong coupling
analysis.

Our analysis of the IIA brane set-ups has been brief as these set-ups are
fairly well studied in the literature. We now turn to describing the bulk
geometry and the associated branes exactly in terms of conformal field theory,
in a suitable limit of the NS$5$ brane background.

\section{Worldsheet Description of the Brane/Orientifold System}
\label{worldsheet}

For type IIA strings on the deformed conifold, one can define a double scaling limit 
\begin{equation}
\mu \rightarrow 0\qquad g_s \rightarrow 0\qquad \text{keeping}\qquad \frac{\mu}{g_s} \qquad \text{fixed}\,. 
\end{equation}
The mass of the D-branes that wrap the three-sphere remains finite.
In this limit, it is conjectured that the decoupled non-gravitational theory
near the singularity has a holographic description as a string theory whose
worldsheet description is the conformal field theory \cite{GiveonKP, GiveonK}
\begin{equation}\label{cigar}
\mathbb{R}^{1,3} \times \left[ \frac{SL(2,\mathbb{R})}{U(1)}\right]_{k=1} \, , 
\end{equation}
where the coset conformal field theory is at supersymmetric level $k=1$. We note in passing that the relation between D-branes in NS$5$ backgrounds and the geometric engineering of gauge theories using branes on the deformed conifold, that was mentioned earlier, is consistent with this conjecture. The topological sub-sector of the CFT \eqref{cigar} has been matched to the topological sector of the deformed conifold theory for both closed strings \cite{Mukhi:1993zb, GhV} and open strings \cite{Ashok:2005xc}.

One can realize SQCD like theories in this set-up by introducing $N_c$
D3-branes localized at the tip of the cigar associated to an $U(N_c)$ gauge
group \cite{FotopoulosNP, AshokMT} and $N_f$ double-sheeted D5-branes filling
the whole of space-time that introduce $N_f$ flavors into the gauge theory
\cite{FotopoulosNP,MurthyT}. We next turn to a detailed discussion of the
crosscaps and boundary states in this background that will allow us to
understand Seiberg duality from a worldsheet perspective.

\subsection{Review of Cylinder Amplitudes and Boundary States}

In \cite{MurthyT} it was argued that a sign flip of the cigar bulk interaction
coefficient $\mu$ leads to an exact conformal field theory realization of
Seiberg duality for $U(N_c)$ gauge theories. We recall that from the cylinder
amplitudes for these branes, one could derive the one-point functions via a
modular bootstrap procedure \cite{Zamolodchikov:2001ah,EguchiS,janbranes}. 
In this way one could associate to characters of
the $N=2$ superconformal algebra (labeled by $J,M$, functions of the
conformal dimension and R-charge) particular Cardy boundary states. A further
crucial ingredient in \cite{MurthyT} was the addition relation:
\begin{equation}\label{DDdiff}
|D5,J=0\rangle = |D3\rangle +\overline{|D5,J=\half\rangle}  \,.
\end{equation}
for exact boundary states. This is a consequence of an $N=2$  superconformal
algebra character identity at level $k=1$, relating the $J=0,M=0$ continuous
character, with the $J=1/2,M=1/2$ continuous character and the identity
character.

We briefly recall a few annulus amplitudes for D-branes in the background 
\eqref{cigar}, as discussed in \cite{EguchiS,janbranes,FotopoulosNP, AshokMT,MurthyT}.
By a $D3$ brane in the background \eqref{cigar}, we mean a D-brane that fills
$\mathbb{R}^{1,3}$ that is tensored with the identity brane in the cigar
direction. We denote the corresponding boundary state as $|D3\rangle$ as above.
Similarly, a $D5$ brane fills $\mathbb{R}^{1,3}$ and is tensored with a brane
labeled by the continuous representation $|D5,J,M\rangle$. We mostly focus on the
branes with $J=M=0$ and $J=M=\half$ and will neglect the $M$ eigenvalue.

The modular bootstrap is implemented by requiring that
 only the identity character appears in the self-overlap of the
$D3$ brane while the continuous character appears in the overlap of the
$D3$ and the $D5$ branes. In what follows, we will use the notation
$q_{o}=e^{-2\pi t}$ and $q_{c}=e^{-2\pi s}$ for the open and closed string
channel moduli. In the open string channel, the
cylinder amplitude for the overlap of the $J=\half$ D$5$ brane with the D$3$ brane
is given by
\begin{multline}\label{JhalfD}
B(D3; J=\half)  = \half\int\frac{dt}{2t} \int \frac{d^4 k}{(2\pi)^4} \frac{e^{-2\pi t k^2} }{\eta(it)^6} \left[\Theta_{1,1}(it)\vartheta_{00}^2(it)+\Theta_{1,1}(it)\vartheta_{01}^2(it) \right. \cr
\left. -\Theta_{0,1}(it)\vartheta_{10}^2(it)\right]\,.
\end{multline}
Similarly for the $J=0$ overlap with the D$3$, we get 
\begin{multline}\label{JzeroD}
B(D3; J=0) =  \half\int\frac{dt}{2t} \int \frac{d^4 k}{(2\pi)^4} \frac{e^{-2\pi t k^2}\, q_{o}^{-1/4} }{\eta(it)^6} \left[\Theta_{0,1}(it)\vartheta_{00}^2(it)-\Theta_{0,1}(it)\vartheta_{01}^2(it) \right. \cr
\left. -\Theta_{1,1}(it)\vartheta_{10}^2(it)\right]\,.
\end{multline}
Using the identities in equation \eqref{thetaid} of the appendix,
one can show that both these
amplitudes vanish. All three boundary states are therefore mutually supersymmetric.

From the identity in \eqref{DDdiff}, we see that from the difference of \eqref{JzeroD} and
\eqref{JhalfD}, one obtains the D$3$ self-overlap. Indeed, the $J=0$
representation of the $N=2$ superconformal algebra is reducible.  Using the modular
bootstrap method, one can modular transform the annulus amplitude into the closed
string channel and read off the D$3$ and D$5$ brane wavefunctions for both
$J=0$ and $J=\half$. This has been done in \cite{EguchiS, janbranes} and we record the NS sector wavefunctions below\footnote{We suppress the $\mu$-dependent phase factor in all wavefunctions (both for the boundary states and the crosscaps to follow). We will come back to this point in Section $4$.}:
\begin{align}\label{Dthreefive}
\Psi_{\sId}^{\sNS,\pm}(P,w) &= \frac{\CC_{\pm}}{\sqrt{2}}\, \frac{\Gamma(\half+iP+\frac{w}{2})\Gamma(\half+iP-\frac{w}{2})}{\Gamma(2iP) \Gamma(1+2iP)}\cr
\Psi^{\sNS,\pm}_{\scont,J=\half}(P,w) &=2\, \frac{\CC_{\pm}}{\sqrt{2}}\, ,\frac{\Gamma(-2iP)\Gamma(1-2iP)}{\Gamma(\half-iP+\frac{w}{2})\Gamma(\half-iP-\frac{w}{2})} \cr
\Psi^{\sNS,\pm}_{\scont,J=0}(P,w) &=2\, \frac{\CC_{\mp}}{\sqrt{2}}\,\cosh2\pi P \,\frac{\Gamma(-2iP)\Gamma(1-2iP)}{\Gamma(\half-iP+\frac{w}{2})\Gamma(\half-iP-\frac{w}{2})}\,,
\end{align}
where $\CC_{+}=1$ and $\CC_{-} = e^{i\pi w}$. Similar expressions also exist
for the RR sectors that can be easily obtained from the expression above
by spectral flow $w\rightarrow w+1$. 

\subsection{M\"obius Amplitudes and Crosscap States}
\label{crosscaps}
To construct crosscap states, we proceed in analogy with the boundary states.
We will perform a modular bootstrap procedure for the M\"obius amplitudes
for the supercoset conformal field theory
at level $k=1$. It is clear how to generalize our analysis to 
other levels, but we will not need the generalization
 for the applications in this paper. Our
starting point will be the requirement that our crosscap states have an
overlap with the identity D$3$-brane which is an $\Omega$-twisted $N=2$
superconformal character (as in \cite{Nakayama}). We then use the open-closed
duality of the M\"obius amplitudes and our knowledge of the D$3$-brane
boundary state to derive corresponding crosscap states. Afterwards, we will
interpret the crosscap states in terms of orientifolds of the bulk superstring
theory by computing the Klein-bottle amplitudes.

Twisting ordinary $N=2$ superconformal characters in the open string channel
by $\Omega$ can be implemented by the substitution  $\tau\rightarrow \tau+\half$ in the
untwisted characters $\chi(\tau)$. One obtains the twisted characters
$\chi_{\Omega}(\tau)$ that take into account the action of $\Omega$ on the
fermionic oscillators. The full M\"obius amplitude that corresponds to the overlap of the crosscap
state labeled by the $J=\half$ continuous character with the D$3$ brane (and vice
versa) is then given by
\begin{multline}\label{JhalfopenC}
M_{\Omega}(D3; J=\half)=  \half\, \int \frac{d^4 k}{(2\pi)^4} \frac{e^{-2\pi t k^2} }{\eta(it+\frac{1}{2})^6} \left[\Theta_{1,1}(it+\half)\big(\vartheta_{00}^2(it+\frac{1}{2})+\vartheta_{01}^2(it+\frac{1}{2})\big) \right.\cr
\left. -\Theta_{0,1}(it+\half)\vartheta_{10}^2(it+\frac{1}{2})\right] \,.
\end{multline}
The M\"obius amplitude for the overlap of the $J=0$ crosscap state with the D$3$ brane can similarly be written as\footnote{The factor of $e^{-\frac{i\pi}{4}}$  ensures that the small $q$ expansion of the characters begins with a real number.} 
\begin{multline}\label{JzeroopenC}
M_{\Omega}(D3; J=0) =\frac{e^{-\frac{i\pi}{4}}}{2}\, \int \frac{d^4 k}{(2\pi)^4} \frac{e^{-2\pi t k^2}\, q_{o}^{-1/4} }{\eta(it+\frac{1}{2})^6} \left[\Theta_{0,1}(it+\half)\big(\vartheta_{00}^2(it+\frac{1}{2})-\vartheta_{01}^2(it+\frac{1}{2})\big) \right.\cr 
\left. -\Theta_{1,1}(it+\half)\vartheta_{10}^2(it+\frac{1}{2})\right]
\end{multline}
Using the same theta function identities that were used to show the vanishing of the
D$5$-D$3$ overlap (for $\tau'=\tau+\half$), we see that both the overlaps
vanish.  Thus both these crosscap states are supersymmetric with respect to
the D$3$ brane.

Similar to the D-brane boundary states, we define the crosscap state corresponding to
the $\Omega$-twisted identity character as the difference between the
$J=0$ and $J=\half$ crosscap state. Let us now execute the modular bootstrap
and determine the closed string one-point functions explicitly.

\subsubsection*{Modular Bootstrap for the Crosscap}

Modular transforming the first term of the $J=\half$ M\"obius amplitude into
the closed string channel,
and using that
the closed string and open string channel parameters are now related by
$s=\frac{1}{4t}$, we get
\begin{align}\label{crosshalfA}
M^{\sNS}_{\Omega}(D3;J=\half)
&= 8e^{\frac{i\pi}{2}}\, \half\int\frac{ds}{128\pi^4}\, \frac{e^{\frac{i\pi}{8}}\,\vartheta_{01}(is+\half)}{\eta(is+\half)^3}\, \int_{0}^{\infty}dP\, 2\, \vartheta_{01}(2is)\, q_{c}^{P^2}\,  \frac{e^{\frac{i\pi}{8}}\,\vartheta_{01}(is+\half)}{\eta(is+\half)^3}
\end{align}
The details of the modular transformation are given in the Appendix. For the
$J=0$ M\"obius amplitude in equation \eqref{JzeroopenC}, 
the $q_{o}^{-\frac{1}{4}}$ factor is first transformed into
\begin{align}
e^{\frac{\pi}{8s}} 
&=  \sqrt{2s}\int_{0}^{\infty}dP\, \cosh \pi P\, e^{-2\pi sP^2} \,.
\end{align}
For the NS sector, we get for the $J=0$ M\"obius amplitude
\begin{align}
M_{\Omega}^{\sNS}(D3; J=0) &= 4\int\frac{ds}{128\pi^4}\, \frac{e^{\frac{i\pi}{8}}\vartheta_{01}(is+\half)}{\eta(is+\half)^3}\, \int_{0}^{\infty}dP\, 2\cosh \pi P\, \vartheta_{10}(2is)\, q_{c}^{P^2}\,  \frac{e^{\frac{i\pi}{8}}\vartheta_{01}(is+\half)}{\eta(is+\half)^3}\,.
\end{align}
Following the modular bootstrap approach that led to the boundary states, one
can use the results for the D-brane wavefunctions in equation \eqref{Dthreefive} to
derive the wavefunctions for the crosscap states. 

Let us begin with the $J=\half$ overlap in  equation \eqref{crosshalfA}: expanding the
theta-function $\vartheta_{01}(2is)$ as we did before, we get
\begin{align}
M^{\sNS}_{\Omega}(D3; J=\half) &= e^{\frac{i\pi}{2}}\,  
\int \frac{ds}{32\pi^4} \, \frac{e^{\frac{i\pi}{8}} \vartheta_{01}(is+\half)}{\eta(is+\half)^3} 
\sum_{n\in \mathbb{Z} } (-1)^{n}\, \int_{0}^{\infty} dP\, 2\, q_{c}^{P^2+n^2}\, \frac{e^{\frac{i\pi}{8}}\vartheta_{01}(is+\half)} {\eta(is+\half)^3}\,.
\end{align}
This should be rewritten in terms of the closed string wavefunctions as
\begin{align}
2\int \frac{ds}{128\pi^4}\frac{e^{\frac{i\pi}{8}}\vartheta_{01}(is+\half)}{\eta(is+\half)^3}\sum_{n\in\mathbb{Z}} \int_{0}^{\infty}\, dP\, \Psi_{\sId}^{\sNS,\mp}(-P,2n)\, \Psi^{\sNS,\pm}_{C,\scont,J=\half}(P,2n)\, ch_{c,\Omega}^{\tilde{\sNS}}(P,2n,is)\,.
\end{align}
Here, the $\Omega$-twisted character $ch_{c,\Omega}^{\tilde{\sNS}}$ is given
by a twisted overlap of the Ishibashi states
\begin{align}
\ch_{c,\Omega}^{\tilde{\sNS}}(P,n,is) & =  q_{c}^{P^2+\frac{n^2}{4}}\, \frac{e^{\frac{i\pi}{8}}\vartheta_{01}(is+\half)} {\eta(is+\half)^3} \,.
\end{align}
Note that in the overlap, only the even winding modes $w=2n$ appear. The crosscap state can therefore be written as a linear combination of Ishibashi states with even winding:
\begin{align}
|C,J=\half;\pm\rangle_{\sNSNS} &= \sum_{n\in \mathbb{Z}}\int_{0}^{\infty}dP\, \Psi_{C,\scont,J=\half}^{\sNS,\pm}(P,2n)\, i^{(L_{0}^{osc}+\bar{L}_{0}^{osc})}\, i^{(J_0^{osc}+\bar{J}_{0}^{osc})}\, |P,2n;\pm\rangle\rangle \,,
\end{align}
where the crosscap wavefunction is given by
\begin{equation}\label{crossWhalf}
\Psi_{C,\scont,J=\half}^{\sNS,\pm}(P,2n) = \frac{8}{\sqrt{2}}\, i\, e^{i\pi n}\, \frac{\Gamma(-2iP) \Gamma(1-2iP)}{\Gamma(\half-iP+n)\Gamma(\half-iP-n)}\,.
\end{equation}
Let us now consider the open string contribution in the RR sector. Similar manipulations yield that the crosscap state is a sum of only those Ishibashi states with odd winding modes
\begin{align}\label{oddishi}
&|C, J=\half;\pm\rangle_{\sRR} = \sum_{n\in \mathbb{Z}}\int_{0}^{\infty}dP\, \Psi_{C,\scont,J=\half}^{\sR,\pm}(P,2n-1)\, i^{(L_{0}^{osc}+\bar{L}_{0}^{osc})}\, i^{(J_0^{osc}+\bar{J}_{0}^{osc})}\, |P,2n\mp1;\pm\rangle\rangle \nonumber\\
&\text{with}\qquad \Psi_{C,\scont,J=\half}^{\sR,\pm}(P,2n\mp1) =  \frac{8}{\sqrt{2}}\, \frac{\Gamma(-2iP) \Gamma(1-2iP)}{\Gamma(\frac{1}{2}-iP+n)\Gamma(\half-iP-n)}\,.
\end{align}

A similar exercise can be carried out for the $J=0$ crosscap state. The wavefunctions take the form
\begin{align}\label{crossWzero}
\Psi_{C,\scont,J=0}^{\sNS,\pm}(P,2n+1) &= \mp\, \frac{8}{\sqrt{2}}\, \cosh\pi P\,  \frac{\Gamma(-2iP) \Gamma(1-2iP)}{\Gamma(1-iP+n)\Gamma(-iP-n)} \nonumber\\
\Psi_{C,\scont,J=0}^{\sR,\pm}(P,2n) &= \mp\, \frac{8}{\sqrt{2}}\, e^{i\pi n}\, \cosh\pi P\,\frac{\Gamma(-2iP) \Gamma(1-2iP)}{\Gamma(1-iP+n)\Gamma(-iP-n)}\,.
\end{align}
Thus, in the NSNS sector, we find that the $J=0$ crosscap has only odd winding modes while the RR sector wavefunctions have even winding modes.

A crosscap state that has a consistent projection in the open string channel on the identity character is the identity crosscap state, whose overlap with the D$3$ gives the twisted identity character. Its wavefunction is just the difference of the $J=0$ and $J=\half$ crosscap states in equations \eqref{crossWzero} and \eqref{crossWhalf}. We will use this crosscap state to engineer our $SO/Sp$ gauge theory in the electric set-up. However, let us first complete the worldsheet analysis and use the crosscap wavefunctions to compute the Klein bottle amplitudes associated to the crosscap states with labels $J=1/2$ and $J=0$.

\subsection{Klein Bottle Amplitudes}

In order to interpret the crosscap states, it is useful to compute the Klein bottle amplitudes. When combined with the torus amplitude, the Klein bottle projects the bulk closed string spectrum. By computing the Klein bottle, we can identity the action of the orientifold projection operator on the closed strings.

Let us consider the $J=\half$ case first. Using the explicit wavefunctions
obtained in the previous section, we get, for the NS$+$NS$+$ overlap
\begin{align}
\cK^{(\half)}_{\sNS+\sNS+}&=\half\int \frac{ds}{128\pi^4}\frac{\vartheta_{00}(is)}{\eta^3(is)}\, \sum_{n\in\mathbb{Z}}\int_{0}^{\infty} dP\, 2^{5}\, \frac{\cosh 2\pi P+\cos 2n\pi}{(\sinh2\pi P)^2}
ch_{c}^{\sNS}(p,n,is)\cr
&= \half\cdot 2^{4}\, \int \frac{ds}{128\pi^4}\, \frac{\vartheta_{00}(is)}{\eta^3(is)}\, \sum_{n\in\mathbb{Z}}\int_{0}^{\infty} dP\, \frac{q_{c}^{P^2+n^2}}{(\sinh \pi P)^2}\frac{\vartheta_{00}(is)}{\eta^3(is)}\cr
&= \half\cdot 2^{4}\, \int \frac{ds}{128\pi^4}\, \frac{\vartheta_{00}(is)}{\eta^3(is)}\,  \int_{0}^{\infty} dP\, \frac{\vartheta_{00}(2is)}{(\sinh \pi P)^2}\, q_{c}^{P^2}\, \frac{\vartheta_{00}(is)}{\eta^3(is)}\,.
\end{align}
Modular transforming to the $t$ channel using $s=\frac{1}{2t}$, we get  
\begin{align}\label{KBNShalfamp}
\cK^{(\half)}_{\sNS+\sNS+} &= \half\int\frac{dt}{128\pi^4\, t^3}\, \frac{\vartheta_{00}(2it)}{\eta^3(2it)}\, \int_{0}^{\infty} \frac{dP}{(\sinh \pi P)^2}\, \left[\Theta_{0,1}(2it)+\Theta_{1,1}(2it)\right]\cr
&\qquad\qquad\qquad\qquad \qquad\qquad\qquad\qquad\qquad
\int_{0}^{\infty} dP'\, \cos2\sqrt{2}\pi PP'\, q_{o}^{P'^2} \, \frac{\vartheta_{00}(2it)}{\eta^3(2it)}\, \cr
&= \half\int\frac{dt}{128\pi^4\, t^3}\, \frac{\vartheta_{00}(2it)}{\eta^3(2it)}\int_{0}^{\infty}  dP' \left[ \rho^{\sNS}_{1,KB} (P') \text{Ch}^{\sNS} (P',0, 2it) \right. \cr
&\left. \qquad\qquad\qquad\qquad\qquad\qquad\qquad\qquad\qquad\qquad+ \rho^{\sNS}_{2,KB} (P') \text{Ch}^{\sNS} (P',1, 2it) \right] \,. 
\end{align}
where we have used the identity
\begin{equation}
\vartheta_{00}(\tau)=\vartheta_{00}(4\tau)+\vartheta_{10}(4\tau) \,.
\end{equation} 
Here, we have defined the extended characters (at level $1$) \cite{EguchiS}
\begin{align}
\text{Ch}^{\sNS}(P,m,\tau) = q^{P^2}\, \Theta_{m,1}(\tau)\, \frac{\vartheta_{00}(\tau)}{\eta^3(\tau)} \,,
\end{align}
and the density of states
\begin{equation}
\label{KBhalfden}
\rho^{\sNS}_{1,KB} (P') =  \rho^{\sNS}_{2,KB} (P') = \int_{0}^{\infty} dP  \frac{\cos2\sqrt{2}\pi PP'}{(\sinh \pi P)^2}\, . 
\end{equation}
We note that the exchange of momentum integrals in $P$ and $P'$ is allowed
 in this case -- there are no subtleties such as those discussed in \cite{Teschner:2000md}. 
 
 One can do similar calculations for the NS$+$NS$-$ and R$+$R$+$ as well. One
 can then check that the total Klein bottle amplitude vanishes. These
 calculations as well as the analogous ones for the $J=0$ crosscap state are
 carried out in Appendix C. One important difference between the $J=0$ and the
 $J=\half$ Klein-bottle computation is that, unlike the infrared divergent density of
 states in \eqref{KBhalfden}, the density of states for the $J=0$ crosscap is
 a $\delta$-function at $P=0$. This point will be crucial later on when we
 engineer the gauge theories of interest and we will come back to this point
 shortly, but we first give a closed string interpretation of these crosscap
 states in terms of projection operators.

\subsection*{Interpretation}
\begin{itemize}
\item Let us first study the bulk superstring theory in which we only add the $J=1/2$
crosscap state. From comparison of the asymptotics (or volume divergence) of
the Klein-bottle amplitude and the torus amplitude, we
conclude that the bulk orientifold is the one obtained from modding out the
type IIB non-critical superstring theory by the operation $\Omega$.
Indeed the Klein bottle amplitude \eqref{KBNShalfamp} has the
interpretation as a trace over the closed string Hilbert space with $\Omega$ inserted.
 
\item If we consider the bulk theory supplemented with the crosscap state
  $J=0$ only, then from the Klein-bottle amplitude, we derive that we have the
  type IIB bulk theory modded out by the operation $\Omega (-1)^n$. The operator
$(-1)^n$ gives an extra minus sign to all Klein-bottle contributions with odd
  momentum $n$ along the angular direction of the cigar, compared to the Klein
bottle amplitude with only $\Omega$ inserted. We note that the
  operation
$(-1)^n$ is coded in coordinate space as the shift $\theta \rightarrow \theta
  + \pi$, which is a symmetry of the cigar winding condensate {\it only} on the
  condition that the parameter $\mu$ is {\em real}. Moreover, geometrically,
  the action only has a fixed point at the tip of the cigar, leading to the
  fact that the corresponding Klein bottle amplitude does not exhibit a volume
  divergence (in the dilaton direction). 
  
\item The $J=\half$ crosscap has a continuous density of states
  \eqref{KBhalfden} which diverges as $P \to 0$ (like the D5 branes) whereas
  the $J=0$ crosscap, as shown in Appendix C, has a $\delta$-function density
  localized strictly at $P=0$. This tells us that the former is extended along
  the cigar direction while the latter is localized at the tip. We can also
  verify this from the form of the wavefunctions in equations \eqref{crossWhalf},
  \eqref{crossWzero}.
  
\item Finally, when we add both crosscap states to the theory, we observe that
  they do not generate cross-terms in the Klein-bottle amplitude, due to the
  orthogonality of the Ishibashi states. When we add both Klein bottle
  amplitudes to the torus amplitude, they generate a combined $(1+\Omega +
  \Omega(-1)^n)$ orientifold/orbifold.  In a closed string theory modded out
  by the above two actions, one would be forced (from the closure of the
  group) to include an orbifold projection $(-1)^n$ as well, leading to a
  projection of the form $\frac{1}{4}(1+\Omega)(1+(-1)^n)$.

\item We note that the $J=1/2$ crosscap states only exchange massive strings,
  including massive gravitons. The exchanged states have worldsheet vertex
operators that differ in conformal dimensions
 by integers. The crosscap state is not charged under massless
RR fields.
\item The $J=0$ crosscap states exchanges massless closed strings, including
  the winding tachyon in the NSNS sector, and the massless RR scalar. 
Again the exchanged states differ by an integer in their worldsheet conformal dimensions. 
\end{itemize}

We move on to finalize our
analysis of amplitudes in the unoriented theory. We compute the overlap of the
crosscap states with the flavour D$5$-branes.

\subsection{D$5$ Brane M\"obius Amplitudes}\label{Dfivemobius}

 We start out by calculating the overlap of the $J=\half$ crosscap with the $J=\half$ D$5$ brane in the NS$+$NS$-$ sector as this corresponds to an NS sector open string amplitude.
  
\begin{multline}
M_{\Omega}^{\sNS}(D5,J=\half; J=\half)= 2\, \int \frac{ds}{128\pi^4}\sum_{n\in \mathbb{Z}} \int_0^{\infty}dP \Psi^{\sNS,\mp}_{\scont,J=\half}(-P,2n)\Psi^{\sNS,\pm}_{C,\scont,J=\half}(P,2n) \cr
q^{P^2+n^2}\left[ \frac{e^{\frac{\pi i}{8}}\vartheta_{01}(is+\half)}{\eta(is+\half)}\right]^2 \,.
\end{multline}
Substituting the wavefunctions using equations \eqref{Dthreefive} and \eqref{crossWhalf}, one gets 
\begin{align}
M_{\Omega}^{\sNS}(D5,J=\half; J=\half)&=4i\int \frac{ds}{128\pi^4}\sum_{n\in\mathbb{Z}}\int_{0}^{\infty}\frac{dP}{(\sinh \pi P)^2}\, e^{i\pi n}\, q_{c}^{P^2+n^2}\left[e^{\frac{i\pi}{8}}\frac{\vartheta_{01}(is+\half)}{\eta^3(is+\half)}\right]^2 \cr
&=4i\int \frac{ds}{128\pi^4}\int_{0}^{\infty}\frac{dP}{(\sinh \pi P)^2}q_{c}^{P^2}\vartheta_{01}(2is)
\left[e^{\frac{i\pi}{8}}\frac{\vartheta_{01}(is+\half)}{\eta^3(is+\half)}\right]^2 \,.
\end{align}
Modular transforming this to the open channel using $s=\frac{1}{4t}$, we obtain 
\begin{align}\label{Mobhalfden}
M_{\Omega}^{\sNS}(D5, J=\half; J=\half)&=\int\frac{dt}{128\pi^4t^3}\frac{e^{\frac{\pi i}{8}}\vartheta_{00}(it+\half)}{\eta^3(it+\half)}\, \int_{0}^{\infty} dP'\, \rho^{\sNS}_{2,M} (P') \,  \text{Ch}^{\sNS}_{\Omega}(P',1,it) \nonumber\\
\text{where}\qquad \text{Ch}^{\sNS}_{\Omega}(P,m,\tau) &=q^{P^2}\, \Theta_{m,1}(\tau)\, 
\frac{e^{\frac{i\pi}{8}} \vartheta_{00}(\tau+\half)}{\eta^3(\tau+\half)} \\
\text{and}\qquad\rho^{NS}_{2,M} & = \int_{0}^{\infty} dP \, \frac{ \cos 2\pi PP'}{(\sinh \pi P)^2} \nonumber\,.
\end{align}
Here we have used the identity
$\vartheta_{10}(\tau+1)=e^{\frac{i\pi}{4}}\vartheta_{10}(\tau)$ and defined
the $\Omega$-twisted extended character $\text{Ch}_{\Omega}\,$. We observe
that this modifies the density of states $\rho_2$ in the D$5$ self overlap
(with a divergent contribution) \cite{EguchiS, MurthyT}. Note that the constants conspire so that in
the $P\rightarrow 0$ limit of the sum of the cylinder and M\"obius amplitudes,
the projection in the open string channel is purely by $\Omega$ on the
massless open string states.  A similar computation yields the characters for
the other sectors Ch$^{\T{\sNS}}_{\Omega}(P',1,it)$ and
Ch$^{\sR}_{\Omega}(P',1,it)$ with the same density of states
\eqref{Mobhalfden}.

Let us repeat this exercise for the overlap of the $J=0$ crosscap with the $J=\half$ D$5$-brane in the NS$+$NS$-$ sector:
\begin{align}
M_{\Omega}^{\sNS}(D5, J=\half; J=0)
&=- 4 \int\frac{ds}{128\pi^4}\int_{0}^{\infty}\frac{dP}{\cosh \pi P}\vartheta_{10}(2is)\, q_c^{P^2}
\left[e^{\frac{i\pi}{8}}\frac{\vartheta_{01}(is+\half)}{\eta^3(is+\half)}\right]^2\,.
\end{align}
Modular transforming this into the open string channel, we get
\begin{align}
\label{Mobzeroden}
M_{\Omega}^{\sNS}(D5, J=\half; J=0)&=\int\frac{dt}{128\pi^4t^3}\frac{e^{\frac{\pi i}{8}}\vartheta_{00}(it+\half)}{\eta^3(it+\half)}\int_{0}^{\infty} dP'\, \rho^{\sNS}_{1,M} (P') \,  \text{Ch}^{\sNS}_{\Omega}(P',0,it), \cr
\rho^{\sNS}_{1,M} & = \int_{0}^{\infty} dP \, \frac{ \cos 2\pi PP'}{(\cosh \pi P)}  \,.
\end{align}
A similar computation yields the characters for the other sectors Ch$^{\T{\sNS}}_{\Omega}(P',0,it)$ and Ch$^{\sR}_{\Omega}(P',0,it)$ with the same density of states \eqref{Mobzeroden}. As shown in Appendix B, where all the amplitudes in the other channels have been computed, the total M\"obius amplitude vanishes in each case:
\begin{equation}
M_{\Omega}^{\sNS}-M_{\Omega}^{\tilde{\sNS}}-M_{\Omega}^{\sR} = 0
\end{equation}
We therefore observe that the boundary states and
crosscap states that we considered are all mutually space-time supersymmetric.
This concludes the calculation of amplitudes that will be relevant for the
gauge theory to be constructed in the next section. These M\"obius amplitudes
will be instrumental in inferring what the global symmetry group is for the gauge
theories we construct in the next section. For now, we conclude this section
with the following observation:
\begin{itemize}
\item The $J=0$ crosscap overlap with the $J=\half$ extended brane gives a
  density of states which is spread a little around $P'=0$ but decays rapidly
  as $P' \to 0$. This is consistent with the fact that the $J=0$ crosscap is localized at
  the tip of the cigar. It is interesting that unlike the previous examples studied for
  localized D-branes, we do not get a sharp $\delta-$function in this case for
  the localized crosscap.
\end{itemize}

\section{Microscopic Description of Seiberg Duality}
\label{flip}

\subsection{The Electric Set-up}

We now have all the ingredients, the boundary states and crosscap states, to engineer the gauge theories of interest to us. We start with the electric description of the gauge theory. To engineer an
$SO(N_c)$ gauge theory with $N_f$ flavours, we consider a configuration of
$N_{c}$ $\ket{D3}$ branes, $N_{f}$ $\ket{D5;J=\half}$ branes and an
orientifold state $\ket{C, \mu}$ which is the difference of the $J=\half$ and $J=0$
crosscap states discussed in the earlier sections
\begin{eqnarray}
\ket{C, \mu} &=& \ket{C,J=M=1/2, \mu} - \ket{C,J=M=0,\mu}.
\end{eqnarray} We first discuss the pure
gauge theory degrees of freedom.

\subsubsection{Pure Gauge Theory}

The orientifold does not act on the gauge field vertex operators themselves,
but has an action on the Chan-Paton factors. Effectively, it makes the $3-3$
strings unoriented.  In the presence of the orientifold, the $D3$ branes
realize a four-dimensional $SO(N_{c})$ gauge theory with $\CN=1$ supersymmetry
on their worldvolume. To see this, one can combine the D3-brane cylinder
amplitude in the oriented theory, divided by $2$ (from the orientifold
projection operator) and multiplied by $N_c$ squared, and the M\"obius
amplitudes of subsection \ref{crosscaps}, divided by $2$, and multiplied by
$N_c$ to obtain the amplitude that codes the open string spectrum projected by
the orientifold operation. The first term in the expansion of the amplitude,
for the $SO(N_c)$ gauge theory is then:
\begin{eqnarray}
Z &=& \frac{N_c^2-N_c}{2} . (2-2) + \dots
\end{eqnarray}
which has the interpretation as counting the transverse polarizations of a vector field in the adjoint of the $SO(N_c)$ gauge group (as well as the corresponding gauginos). The $Sp$ gauge group is realized by taking the opposite (overall) sign of the orientifold plane.

We can confirm the above picture as follows.
From the ratio of one-point functions for the RR scalar, 
we can read of the charge
  of the crosscap state relative to the charge of the D5-brane. A short
  calculation gives that the ratio of the $J=0$ crosscap charge to that of the
  D5-brane is $-4$. It is therefore $+2$ times the charge of an unmirrored
  D3-brane. Therefore minus the $J=0$ crosscap state projects onto $SO$
  gauge theories on the D3-brane.

We note in passing that for the gauge theory without flavours, we can argue 
following \cite{AshokMT} that the breaking of the anomalous $U(1)_R$ is
encoded in the backreaction of the D3-branes {\em and} the orientifold plane
on the RR-scalar. The scalar will develop a dependence on the angular cigar
variable that is proportional to $N_c \mp 2$ for $SO/Sp$ gauge theories. Taken
into account the properly quantized shift symmetry of the RR-scalar, this
then codes the breaking of $U(1)_R$ to $Z_{2 \check{h} }$ in the quantum theory (where
$\check{h}$ is the dual coxeter number of the gauge algebra). 

\subsubsection{Adding Flavour}

We now turn to adding flavour. The $J=\half$ brane does not have any massless localized modes in an overlap with either of the crosscaps. This may be seen from the small $q$ expansions of the M\"obius overlaps in Section \ref{Dfivemobius} and in the Appendix B. Thus the only other massless four-dimensional modes in this setup arise from the unoriented $D3-D5$ strings.
These are chiral superfields $Q^{ia}, a=1,\ldots, N_{c}, i = 1,\ldots, N_{f}$
which fall into the vector representation of $SO(N_{c})$.  The theory on the
worldvolume of the branes at low energies is thus a four-dimensional
$SO(N_{c})$ gauge theory with $N_{f}$ quarks, which has $\CN=1$
supersymmetry. 

From the D$5$ M\"obius amplitudes, we also read off that the global symmetry
group is indeed $Sp(N_{f})$. This can be seen as follows: while the massless D3-brane spectrum is projected by both the $J=0$ and the $J=1/2$ crosscap, the D5-brane massless spectrum is
only affected in the open string channel by the non-compact $J=1/2$ crosscap.
This can be checked by expanding the M\"obius amplitudes for the D5-brane with
the $J=0$ crosscap and finding only massive modes in the open string channel
(see equations (\ref{JzerocrossD5half1}, \ref{JzerocrossD5half2},
\ref{JzerocrossD5half3})). Moreover, by expanding the M\"obius amplitudes that
project the massless modes in the open string channel for the D3-brane (see
subsection \ref{crossD3}) compared to the D5-brane (see subsection
\ref{crossD5}), we see that the M\"obius amplitudes have a relative minus
sign. This can be traced back to the relative sign in the full orientifold
that corresponds to subtraction of the $J=0$ and the $J=1/2$ crosscaps (plus
the fact that the main contribution to the D3-brane M\"obius amplitude is from
the $J=0$ crosscap). The relative minus sign between the D$3$ and D$5$
M\"obius amplitudes codes the fact that the global symmetry group is $Sp$ when
the gauge group is $SO$ (and vice versa). We believe this to be a relevant check on our brane set-up. 

\subsection{Worldsheet Analysis of $\mu-$Transitions and the Magnetic Theory}

Now that we have identified the relevant boundary states and crosscaps, we can
proceed, following \cite{MurthyT}, to infer Seiberg duality from the
worldsheet. In what follows, we use the known transformations of the D-brane
states under the $\mu$-flip transformation $\mu\rightarrow -\mu$, the
consequent rearrangement of the D-branes in the charge lattice, mutual
supersymmetry with the final $\mu$-flipped configuration as well as some
physical statements about tensions  to infer the behaviour of the
crosscap states under the $\mu-$flip. We first review the transformation of
the boundary states following.  Since most of the details can be found in \cite{MurthyT}, we will be brief in  our discussion.

\subsubsection{Behaviour of Boundary States and Crosscaps under $\mu-$Transitions}

The behavior of the D3 and D5-brane boundary states under a sign-flip
 of the coefficient $\mu$ of the 
cigar interaction term (i.e. the N=2 Liouville interaction term corresponding
to a winding condensate) was determined in \cite{MurthyT}. The first ingredient
in the derivation is the fact that the bulk vertex operators
satisfy a reflection relation, and therefore their one-point functions
satisfy it as well \cite{FateevZZ}. This determines the $\mu$ dependence of the one-point
functions for the boundary states, as well as the crosscap states. The
 $\mu$-dependence of the closed string one-point functions in the NSNS sector
amounts to an overall factor of $\mu^{iP-Q/2} \bar{\mu}^{iP+Q/2}$ where $P$ is
 the momentum of the closed string, and $Q$ is the (worldsheet left-moving)
 $U(1)_R$ charge of the
 closed string vertex operator.

We recall that D3-branes are localized objects with a well-defined mass. The
gravitational backreaction is dominantly on the on-shell mode with asymptotic
behaviour $e^{-2\rho}$, $\rho$ being the radial direction of the cigar
\cite{AshokMT}. Under $\mu \rightarrow -\mu$, the one-point function for this
mode flips sign. That shows that the NSNS part of the D3-brane boundary state
needs to pick up an explicit minus sign, when we demand that it maps into a
physical D-brane with positive mass \cite{MurthyT}.

For the NSNS part of the D5-brane boundary state, the dominant backreaction is
onto the cosmological constant, which does not pick up a minus sign under the
$\mu \rightarrow -\mu$ transformation. Moreover, since the D5-brane changes
orientation under the $\mathbb{Z}_2$ transformation \cite{MurthyT}, the RR
part of the boundary state flips sign. Charge conservation then fixes that the
D3-brane RR boundary state flips sign as well (leading to a simple overall
minus sign in the boundary state).

Thus, under the $\mathbb{Z}_2$ transformation, we have the following transformation of the boundary states \cite{MurthyT}:
\begin{align}
N_c\, |D3, \mu \rangle & \longrightarrow N_c\, (-| D3, - \mu \rangle) \cr
N_f\, \overline{ |D5, J=1/2=M , \mu \rangle } &  \longrightarrow N_f\, |D5, J=1/2=M,-\mu \rangle \cr
&\longrightarrow N_f\, \overline{|D5, J=M=0, -\mu\rangle}+N_f\, (-\overline{|D3,-\mu\rangle}) \,.
\end{align}
After the annihilation of the $N_c$ D3-branes with $N_f$ anti-D3-branes (for
$N_f \ge N_c-2$), we remain with $N_f-N_c$ color anti-branes, as well as $N_f$
mutually supersymmetric flavor branes (at $J=0$). Thus, on the magnetic side,
the colour gauge theory is realized on $N_f-N_c$ branes of the form
$(-\overline{|D3,-\mu\rangle})$.

We now turn to the determination of the behaviour of the crosscap states under
the $\mathbb{Z}_2$ transformation. Our attitude will be slightly different
than
in the D-brane case. We will only attempt to describe the final configuration
that results after the system has fully relaxed to the supersymmetric
configuration.
To that end, we would like to understand the crosscaps
which are supersymmetric with respect to the final configuration of D-branes
above in the magnetic theory at $-\mu$. Mutual supersymmetry with the branes
$(-\overline{|D3,-\mu\rangle})$ and $ \overline{|D5, J=M=0, -\mu\rangle}$
dictates that the relative sign of the NSNS and RR parts of the crosscap
states change (as for the D-branes) when we go from $\mu \to -\mu$.
Moreoever from the pure bulk theory before double scaling, we know that 
the exchange of the NS5 and NS5' branes before any D-branes are present is
a trivial operation. That remains true in the presence of the orientifold
plane, so the Klein-bottle amplitude does not change under the $\mu$-flip.

We further expect that the gauge theory which is realized on the localized branes is projected onto an SO group as before. This projection arises from a  M\"obius amplitude which is an overlap of the new crosscap with 
 the localized branes in the magnetic theory $(-\overline {|D3,-\mu \rangle} )$. These conditions fix the transformation of the crosscap:
\begin{align}
|C, \mu \rangle & \longrightarrow  -\overline{|C, \mu \rangle} \, .
\end{align}
We can further argue for this transformation rule by observing that
the tension of the $J=0$ crosscap is measured by its coupling to the tachyon
winding mode and is proportional to $\mu$. Like the localized D-brane, this
also changes sign with $\mu$. That leads to the explicit minus sign in the
transformation rule.\footnote{We should note that it is not obvious in this formalism that the tension of
the magnetic orientifold plane in the asymptotic region, 
as measured in the overlap with the magnetic
flavour brane remains positive (as appropriate for an orientifold plane that
projects onto an $Sp$ global symmetry group). This is hidden in the fact that
the propagator appearing in the calculation of the overlap is $\mu$
dependent.}

The charge of the final state is opposite to the initial state when measured correctly with the new axion which has also changed sign. Unlike the D-brane case, we cannot argue at the level
of boundary/crosscap states that there is a rearrangment of the basis in the
final state, since charge is actually not conserved in this situation
described above. The only possibility consistent with charge conservation, as
we shall consider below, is that there are new charge-carrying localized
states which are created during the transition. These will turn out to be D$3$
branes.

Intuitively speaking, the localized part of the orientifold, namely the $J=0$
crosscap, accounts for the charge difference between the orientifold
four-planes that end on a given NS5-brane from opposite direction in the type
IIA set-up. We will see this more precisely in the next section where we
exhibit pairs of Seiberg dual gauge theories. It will turn out that the axionic
charge measured by the RR scalar in the worldsheet description keeps track of
the spacetime linking number in the IIA set-up.

Finally, we note that the $\mathbb{Z}_2$ flip on the parameter $\mu$ can be
generalized to a transformation of the parameter $\mu$ along the {\em real
  line}. (The parameter $\mu$ needs to be real in order for the $\Omega
(-1)^n$ orientifold to exist. Another way to see this is to observe that there
is no Fayet-Iliopoulos parameter
to turn on  for the $SO/Sp$ gauge theories.)  Thus, in
contrast to the non-critical set-up for $SU$ gauge theories, we necessarily go
through a region of strong coupling. This is the basic reason why we here take the
attitude
to describe the final orientifold configuration from scratch, without attempting to follow
it closely through the whole transition process.  As shown in \cite{EvansJS}, it is this
motion through the strong coupling region that creates of the extra localized
charged D-brane states, which leads to a conservation of the axionic charge in
the process.

\subsubsection{Seiberg Duality}

It will be useful to be explicit about how charge is conserved under the $\mathbb{Z}_2$
transformation. First of all, in the initial configuration, we measure
axionic charge with respect to the massless RR scalar. In the magnetic
theory, we will measure the axionic charge with respect to a sign-flipped RR
scalar (since it flips sign under the $\mathbb{Z}_2$ transformation). In
the electric theory we have $N_c$ (unmirrored) color D3-branes and
$N_f$ flavor anti-D5 (J=1/2) branes, giving rise to an $SO(N_c)$ gauge theory
with $Sp(N_f)$ flavour group.

Recalling that in the oriented theory, the axionic charge of the D$5$ is
$\half$ that of the D$3$ \cite{MurthyT}, this configuration contributes to the axionic charge
$N_c-\half N_f$. From the previous section, we conclude that the
orientifold plane contributes $-2$ to the charge, for a total axionic charge
of $N_c-\half N_f-2$. As mentioned earlier, we observe that this coincides
with the linking number $l_{NS}$ defined
for the NS brane in the type IIA set-up in \eqref{linking}. The charge also
measures the anomaly in a specific linear combination of the axial flavour
$U(1)_A$ charge and the $U(1)_R$ charge (which is proportional to $Q_R+ 1/2 Q_A$ in units where
the R-charge of the gaugino and the axial charge of a left-handed quark are both equal to one).

In the magnetic description, after the annihilation of the $N_c$ D3-branes with $N_f$ anti-D3-branes
(for $N_f \ge N_c-2$), we remain with $N_f-N_c$ color anti-branes, as well as
$N_f$ mutually supersymmetric $J=0$ flavor branes. Furthermore we have a
crosscap state $-\overline{\ket{C;-\mu}}$ which gives a total charge of $N_{c} -
\half N_{f} +2$.

The axionic charge in the electric and magnetic configurations differ by four (in the units of
$D3$ brane charge). Charge conservation therefore necessarily implies that
four $D3$ branes must be created while passing through the strong-coupling
region at $\mu=0$, just as in the discussion of Seiberg duality in the type
IIA NS$5$ brane set up \cite{ElitzurGK}. Thus the gauge group in the magnetic
theory is $SO(N_f-N_c + 4 )$. As in the electric setup, there are no new
massless modes being created by the orientifold. Similarly, for the same
reasons as in the electric set-up, the global symmetry group is $Sp(N_{f})$ also in the magnetic configuration.

The other massless localized modes in the magnetic setup are the magnetic quarks arising from the $D3-D5$ strings and the meson from the $D5-D5$ overlap. As in the $SU(N)$ case \cite{MurthyT}, the overlap of the extended brane at $J=0$ gives rise, in addition to a six-dimensional continuum of modes, to a {\em four dimensional localized} massless scalar with the quantum numbers of the meson. This arises from a correct convergent definition of the overlap which involves a contour prescription \cite{Teschner:2000md,MurthyT}\footnote{This is the only mode that arises in the localized contribution of the $D5$ overlap. There is another massless localized mode with the quantum numbers $A^{ia}$ which arises in the $D3-D5(J=0)$ overlap, but the symmetries forbid a minimal gauge coupling of this mode with any of the massless fields. This is interpreted as the fact that this gauge field gets a mass, as suggested by the ten-dimensional picture \cite{HananyW}.}.

To summarize, the magnetic theory has gauge fields, quarks, as well as mesonic degrees of freedom. As in the electric setup, the strings fluctuations which gave rise to the quarks and mesons are now also unoriented. The matter content of this gauge theory is $N_{f}$ quarks $q^{ia}, a=1,\ldots,
(N_{f}-N_{c}), i=1,\ldots, N_{f}$ in the vector representation and a meson
$M^{ij}$ in the symmetric representation of the $Sp(N_{f})$ global flavor
symmetry.

\section{Conclusions}\label{conclusions}

The main technical result in our work is the construction of exact crosscap states in the cigar background 
using the modular bootstrap method. Using these states in conjunction with the known
D-brane boundary states \cite{FotopoulosNP, AshokMT} in this background, we
engineered four dimensional $SO/Sp$ gauge theories with fundamental flavours.
We found that microscopically, we could interpret Seiberg duality as a
re-arrangement of the basis of boundary states and crosscaps under the
$\mu\rightarrow -\mu$ transformation in the conformal field theory moduli 
space.

The exact microscopic realization of Seiberg duality in ${\cal N} =1$ $SO/Sp$
gauge theories with flavour is rendered more subtle than its $SU(N_c)$
counterpart 
\cite{MurthyT} by the fact that the $\mu-$transition necessarily takes us
through the strong coupling region. In the exact realization this is expressed
through the fact that extra boundary states need to be created while going
through the strong coupling region of the gauge theory. This was derived in
the non-critical set-up by using a charge conservation argument similar to the
one used in the context of the IIA brane set-ups we discussed in the
introduction. The relevant charge in this case turns out to be the axionic
charge measured by the RR scalar.

The crosscap states that we used to engineer the gauge theory with flavour
still have a geometric interpretation in this highly curved regime of string theory.
In principle, one would like to connect these orientifold crosscap states to
their asymptotically flat ancestors, tracking them through the double scaling
limit \cite{GiveonK}, thereby connecting them to geometric orientifold planes in flat space.
One can use the added control over these states afforded by the CFT to probe 
in detail the gauge theory on the branes beyond the information which can be read off from the 
singular ten-dimensional setup. 
As discussed in \cite{MurthyT},  the full theory on the branes at non-zero energies could have contributions from other open and closed string modes. Regarding this point, we note that in the 
presence of the orientifolds we must also take into account any potential contributions 
from the closed string twisted sectors arising from the various Klein bottles. 

We only discussed boundary states and crosscaps based on the continuous
representations of the $\CN=2$ superconformal algebra. These are
double-sheeted objects on the cigar and as a result, there were no constraints
coming from tadpole cancellation. However, it is probable that including
boundary states and crosscaps based on discrete representations would allow us
to include chiral matter into the gauge theory through clever cancellation of
RR tadpoles arising from D-branes and orientifold planes. We leave this as an open
problem.

Finally, it would be very interesting to generalize the implementation of
Seiberg duality in non-critical strings to more general gauge theories with
product gauge groups and bi-fundamental matter obtained by putting
supersymmetric branes in (oriented or unoriented) non-compact Gepner models.
Such theories have been studied in the context of geometric engineering in
\cite{CachazoFIKV}. It would be interesting to see if the description of
Seiberg duality given in that reference (for instance, as Weyl reflections of
the simple roots in the special case of the ADE quivers) can be given a
worldsheet description. When the moduli of the conformal field theory
are complex, we expect Seiberg duality to emerge as a monodromy in the space of couplings. When orientifolds are present, based on the simple example studied here, we expect that the duality will manifest itself as a rearrangement of boundary states and crosscaps under discrete actions on  the space of couplings of suitable Gepner (boundary) conformal field theories.

\section*{Acknowledgements}

We would like to thank Eleonora Dell'Aquila, Jaume Gomis, Amihay Hanany,
Kentaro Hori, Vasilis Niarchos and Marco Serone for useful discussions. The
research of SA at Perimeter Institute for Theoretical Physics is supported in
part by the Government of Canada through NSERC and by the Province of Ontario
through MRI. SA would also like to thank the hospitality of the ICTP at
Trieste, where part of this research was carried out. SM would like to
acknowledge the hospitality of the high energy group at the Perimeter
Institute where part of this work was carried out.  The research of JT was
partially supported by the RTN European programme MRTN-CT2004-005104.

\begin{appendix}

\section{Conventions}\label{conv}

Our definitions for the 
$\vartheta_{ab}$ functions are:
\begin{align}
\vartheta_{00}(\tau,\nu)&=\sum_{n=-\infty}^{\infty}q^{n^2/2}\, z^n  \cr
\vartheta_{01}(\tau,\nu)&=\sum_{n=-\infty}^{\infty}(-1)^n\, q^{n^2/2}\, z^n  \cr
\vartheta_{10}(\tau,\nu)&=\sum_{n=-\infty}^{\infty}q^{(n-1)^2/2}z^{n-1/2}  \cr
\vartheta_{11}(\tau,\nu)&=-i\sum_{n=-\infty}^{\infty}(-1)^n\, q^{(n-1)^2/2}z^{n-1/2}  \,,
\end{align}
where $q=e^{2\pi i \tau}$ and $z=e^{2 \pi i \nu}$. 
We also define the level $k$ $\Theta$ functions:
\begin{align}
\Theta_{m,k}(\tau,\nu)=\sum_{n=-\infty}^{\infty}q^{k(n+\frac{m}{2k})^2}\, z^{k(n+\frac{m}{2k})} \,.
\end{align}
We will only use the level $k=1$ $\Theta_{m,1}(\tau,\nu)$ functions with
$m=0,1$. The level $1$ $\Theta_{m,1}(\tau,\nu)$ functions are related to the $\vartheta(\tau,\nu)$ functions as follows:
\begin{align}
\Theta_{0,1}(\tau,\nu)=\vartheta_{00}(2\tau,\nu) \qquad \Theta_{1,1}(\tau,\nu)=\vartheta_{10}(2\tau,\nu)\,.
\end{align}
The following identities are handy in verifying space-time supersymmetry:
\begin{align}\label{thetaid}
\vartheta_{00}^2(\tau) &= \vartheta_{00}^2(2\tau)+ \vartheta_{10}^2(2\tau) \cr
\vartheta_{01}^2(\tau) &= \vartheta_{00}^2(2\tau)- \vartheta_{10}^2(2\tau) \cr
\vartheta_{10}^2(\tau) &=2\, \vartheta_{00}(2\tau)\vartheta_{10}(2\tau) \,.
\end{align}
Apart from the usual modular transformation rules of the $\vartheta$-functions, the following particular  modular transformations are useful in the unoriented theory \cite{Nakayama}:
\begin{align}\label{orient}
\vartheta_{00}(\frac{i}{4s}+\half) &=\sqrt{2s}\, e^{\frac{i\pi}{4}}\, \vartheta_{01}(is+\half) \cr
\vartheta_{01}(\frac{i}{4s}+\half) &=\sqrt{2s}\, e^{-\frac{i\pi}{4}}\, \vartheta_{00}(is+\half)\cr
\vartheta_{10}(\frac{i}{4s}+\half) &=\sqrt{2s}\, \vartheta_{10}(is+\half)\cr
\eta(\frac{i}{4s}+\half) &= \sqrt{2s}\, \eta(is+\half) \, .
\end{align}

\section{Modular Transformation of M\"obius Amplitudes}

\subsection{Overlaps of crosscaps with D$3$ branes} \label{crossD3}

Let us consider equation \eqref{JhalfopenC} and modular transform the
expression into the closed string channel. Performing the momentum integration, we get
\begin{equation}
\int \frac{d^4k}{16\pi^4} e^{2\pi i\tau k^2} = -\frac{1}{16\pi^4}\, \frac{1}{4\tau^2} = \frac{s^2}{4\pi^2} \,.
\end{equation}
Let us consider the first term in equation \eqref{JhalfopenC}: using the
identities in \eqref{thetaid}\ the integral over the modular parameter can be
written as
\begin{align}
M^{\sNS}_{\Omega}(D3; J=\half) =  \int ds\frac{s}{8\pi^4}\, 
\frac{\vartheta_{00}(\frac{i}{4s}+\half)}{\eta(\frac{i}{4s}+\half)^3}\, e^{\frac{i\pi}{4}}\, \vartheta_{10}(\frac{i}{2s})\,  
\frac{\vartheta_{00}(\frac{i}{4s}+\half)}{\eta(\frac{i}{4s}+\half)^3} \,.
\end{align}
Using the modular transformations \eqref{orient}\ this can be written in the closed string channel as
\begin{align}\label{crosshalfone}
M^{\sNS}_{\Omega}(D3; J=\half) &= e^{\frac{i\pi}{4}}\, \int\, ds\, \frac{s}{8\pi^4}\, 
\left [\frac{1}{2s}\, \frac{e^{\frac{i\pi}{4}}\,\vartheta_{01}(is+\half)}{\eta(is+\half)^3} \right]
\left[\frac{1}{\sqrt{2s}}\frac{e^{\frac{i\pi}{4}}\, \vartheta_{01}(is+\half)\vartheta_{01}(2is)}{\eta(is+\half)^3} \right] \cr
&=e^{\frac{i\pi}{2}}\, \int\, \frac{ds}{16\sqrt{2s}\, \pi^4}\,  \frac{e^{\frac{i\pi}{8}}\,\vartheta_{01}(is+\half)}{\eta(is+\half)^3}\, \vartheta_{01}(2is)\,\frac{e^{\frac{i\pi}{8}}\,\vartheta_{01}(is+\half)}{\eta(is+\half)^3} \cr
&= 8e^{\frac{i\pi}{2}}\, \int\frac{ds}{128\pi^4}\, \frac{e^{\frac{i\pi}{8}}\,\vartheta_{01}(is+\half)}{\eta(is+\half)^3}\, \int_{0}^{\infty}dP\, 2\, \vartheta_{01}(2is)\, q_{c}^{P^2}\,  \frac{e^{\frac{i\pi}{8}}\,\vartheta_{01}(is+\half)}{\eta(is+\half)^3}\,.
\end{align}
where $q_{c}=e^{-2\pi s}$ as usual. Similar manipulations for the $\tilde{\sNS}$ and $\sR$ sector terms in \eqref{JhalfD} yield, after modular transformation,
\begin{align}\label{crosshalfB}
M^{\tilde{\sNS}}_{\Omega}(D3; J=\half) &= -8e^{-\frac{i\pi}{2}}\, \int\frac{ds}{128\pi^4}\, 
\frac{e^{\frac{i\pi}{8}}\vartheta_{00}(is+\half)}{\eta(is+\half)^3}\, \int_{0}^{\infty}dP\, 2\, \vartheta_{01}(2is)\, q_{c}^{P^2}\,  \frac{e^{\frac{i\pi}{8}}\vartheta_{00}(is+\half)}{\eta(is+\half)^3}\cr
M^{\sR}_{\Omega}(D3; J=\half) &= 8\int\frac{ds}{128\pi^4}\, \frac{\vartheta_{10}(is+\half)}{\eta(is+\half)^3}\, \int_{0}^{\infty}dP\, 2\, \vartheta_{10}(2is)\, q_{c}^{P^2}\,  \frac{\vartheta_{10}(is+\half)}{\eta(is+\half)^3}\,.
\end{align}
The $J=0$ M\"obius amplitudes can be similarly modular transformed as follows:
\begin{align}
M_{\Omega}^{\sNS}(D3;J=0) &= 8\int\frac{ds}{128\pi^4}\, \frac{e^{\frac{i\pi}{8}}\vartheta_{01}(is+\half)}{\eta(is+\half)^3}\, \int_{0}^{\infty}dP\, 2\cosh \pi P\, \vartheta_{10}(2is)\, q_{c}^{P^2}\,  \frac{e^{\frac{i\pi}{8}}\vartheta_{01}(is+\half)}{\eta(is+\half)^3}\cr
M_{\Omega}^{\tilde{\sNS}}(D3; J=0) &= -8\int\frac{ds}{128\pi^4}\, \frac{e^{\frac{i\pi}{8}}\vartheta_{00}(is+\half)}{\eta(is+\half)^3}\, \int_{0}^{\infty}dP\, 2\cosh \pi P\, \vartheta_{10}(2is)\, q_{c}^{P^2}\,  \frac{e^{\frac{i\pi}{8}}\vartheta_{00}(is+\half)}{\eta(is+\half)^3}\cr
M_{\Omega}^{\sR}(D3; J=0) &= 8\int\frac{ds}{128\pi^4}\, \frac{\vartheta_{10}(is+\half)}{\eta(is+\half)^3}\, \int_{0}^{\infty}dP\, 2\cosh \pi P\, \vartheta_{01}(2is)\, q_{c}^{P^2}\,  \frac{\vartheta_{10}(is+\half)}{\eta(is+\half)^3} \,.
\end{align}
One can check that the total M\"obius amplitude vanishes:
\begin{equation}
M_{\Omega}^{\sNS} - M_{\Omega}^{\T\sNS} -M_{\Omega}^{\sR}  = 0
\end{equation}

\subsection{Overlaps of Crosscaps with D$5$ Branes}
\label{crossD5}
In this section, we merely record the overlaps of the $J=\half$ and $J=0$ crosscap states with the $J=\half$ and $J=0$ D-branes in the open string channel. These turn to be important to infer the flavour gauge group in the unoriented theory. 

\subsubsection*{Overlaps of $J=\half$ D$5$ brane with $J=\half$ crosscap}

\begin{align}
M_{\Omega}^{\sNS}&=\int\frac{dt}{128\pi^4t^3}\int_{0}^{\infty}\frac{dP}{(\sinh \pi P)^2} \int_{0}^{\infty}dP'\, \cos 2\pi PP' q_o^{P'^2} \Theta_{1,1}(it+\half)\, \left[\frac{\vartheta_{00}(it+\half)}{\eta^3(it+\half)}\right]^2 \,.
\end{align}
\begin{align}
M_{\Omega}^{\tilde{\sNS}}&=-\int\frac{dt}{128\pi^4t^3}\int_{0}^{\infty}\frac{dP}{(\sinh \pi P)^2} \int_{0}^{\infty}dP'\, \cos 2\pi PP' q_o^{P'^2} \Theta_{1,1}(it+\half)\, \left[\frac{\vartheta_{01}(it+\half)}{\eta^3(it+\half)}\right]^2 \,.
\end{align}
\begin{align}
M_{\Omega}^{\sR}&=\int\frac{dt}{128\pi^4t^3}\int_{0}^{\infty}\frac{dP}{(\sinh \pi P)^2} \int_{0}^{\infty}dP'\, \cos 2\pi PP' q_o^{P'^2} \Theta_{0,1}(it+\half)\, \left[\frac{\vartheta_{10}(it+\half)}{\eta^3(it+\half)}\right]^2 \,.
\end{align}

\subsubsection*{Overlaps of $J=\half$ D$5$ brane with $J=0$ crosscap}

\begin{align}
  M_{\Omega}^{\sNS}&=e^{\frac{i\pi}{2}}
  \int\frac{dt}{128\pi^4t^3}\int_{0}^{\infty}\frac{dP}{\cosh\pi
    P}\int_{0}^{\infty}dP' \cos2\pi
  PP'q_{o}^{P'^2}\Theta_{0,1}(it+\half)\left[e^{\frac{i\pi}{8}}\frac{\vartheta_{00}(it+\half)}{\eta^3(it+\half)}\right]^2\,.
\label{JzerocrossD5half1}
\end{align}
\begin{align}
  M_{\Omega}^{\tilde{\sNS}}&=e^{\frac{i\pi}{2}}
  \int\frac{dt}{128\pi^4t^3}\int_{0}^{\infty}\frac{dP}{\cosh\pi
    P}\int_{0}^{\infty}dP' \cos2\pi
  PP'q_{o}^{P'^2}\Theta_{0,1}(it+\half)\left[e^{\frac{i\pi}{8}}\frac{\vartheta_{01}(it+\half)}{\eta^3(it+\half)}\right]^2\,.
\label{JzerocrossD5half2}
\end{align}
\begin{align}
M_{\Omega}^{\sR}&=-e^{-\frac{i\pi}{4}}
\int\frac{dt}{128\pi^4t^3}\int_{0}^{\infty}\frac{dP}{\cosh\pi
  P}\int_{0}^{\infty}dP' \cos2\pi
PP'q_{o}^{P'^2}\Theta_{1,1}(it+\half)\left[\frac{\vartheta_{10}(it+\half)}{\eta^3(it+\half)}\right]^2\,.
\label{JzerocrossD5half3}
\end{align}

\subsubsection*{Overlaps of $J=0$ D$5$ brane with $J=\half$ crosscap}

\begin{align}
  M_{\Omega}^{\sNS}&=
  \int\frac{dt}{128\pi^4t^3}\int_{0}^{\infty}\frac{\cosh\pi PdP}{(\sinh\pi
    P)^2}\int_{0}^{\infty}dP' \cos2\pi
  PP'q_{o}^{P'^2}\Theta_{1,1}(it+\half)\left[\frac{\vartheta_{00}(it+\half)}{\eta^3(it+\half)}\right]^2\,.
\end{align}
\begin{align}
M_{\Omega}^{\tilde{\sNS}}&= -\int\frac{dt}{128\pi^4t^3}\int_{0}^{\infty}\frac{\cosh\pi PdP}{(\sinh\pi P)^2} \int_{0}^{\infty}dP'\, \cos 2\pi PP' q_o^{P'^2} \Theta_{1,1}(it+\half)\, \left[\frac{\vartheta_{01}(it+\half)}{\eta^3(it+\half)}\right]^2 \,.
\end{align}
\begin{align}
M_{\Omega}^{\sR}&=\int\frac{dt}{128\pi^4t^3}\int_{0}^{\infty}\frac{\cosh\pi PdP}{(\sinh \pi P)^2} \int_{0}^{\infty}dP'\, \cos 2\pi PP' q_o^{P'^2} \Theta_{0,1}(it+\half)\, \left[\frac{\vartheta_{10}(it+\half)}{\eta^3(it+\half)}\right]^2 \,.
\end{align}

\subsubsection*{Overlaps of $J=0$ D$5$ brane with $J=0$ crosscap}

\begin{align}
M_{\Omega}^{\sNS}&= -e^{\frac{i\pi}{2}}\int\frac{dt}{128\pi^4t^3}\int_{0}^{\infty}dP\int_{0}^{\infty}dP' \cos2\pi PP'q_{o}^{P'^2}\Theta_{0,1}(it+\half)\left[e^{\frac{i\pi}{8}}\frac{\vartheta_{00}(it+\half)}{\eta^3(it+\half)}\right]^2\,.
\end{align}
\begin{align}
M_{\Omega}^{\tilde{\sNS}}&=-e^{\frac{i\pi}{2}}\int\frac{dt}{128\pi^4t^3}\int_{0}^{\infty}dP\int_{0}^{\infty}dP' \cos2\pi PP'q_{o}^{P'^2}\Theta_{0,1}(it+\half)\left[e^{\frac{i\pi}{8}}\frac{\vartheta_{01}(it+\half)}{\eta^3(it+\half)}\right]^2\,.
\end{align}
\begin{align}
M_{\Omega}^{\sR}&=e^{-\frac{i\pi}{4}}\int\frac{dt}{128\pi^4t^3}\int_{0}^{\infty}dP\int_{0}^{\infty}dP' \cos2\pi PP'q_{o}^{P'^2}\Theta_{1,1}(it+\half)\left[\frac{\vartheta_{10}(it+\half)}{\eta^3(it+\half)}\right]^2\,.
\end{align}

In both cases, one can check that
\begin{equation}
M_{\Omega}^{\sNS}-M_{\Omega}^{\T{\sNS}}-M_{\Omega}^{\sR}=0 \,.
\end{equation}

\section{Klein Bottle Amplitudes}

In this appendix, we collect some technical details about the Klein botle amplitudes.
The computations of the NS$+$NS$-$ and R$+$R$+$ contributions for the
$J=\half$ crosscap state leads to the expressions:
\begin{multline}
\cK^{(\half)}_{\sNS+\sNS-}= \half\, 2^4\,\int \frac{ds}{128\pi^4}\frac{\vartheta_{01}(is)}{\eta^3(is)}\,  \int_{0}^{\infty} dP\, \frac{\vartheta_{00}(2is)}{(\sinh \pi P)^2}\, q_{c}^{P^2}\, \frac{\vartheta_{01}(is)}{\eta^3(is)}\cr
= \half\, \int\frac{dt}{128\pi^4\, t^3}\, \frac{\vartheta_{10}(2it)}{\eta^3(2it)}\, \int_{0}^{\infty} \frac{dP}{(\sinh \pi P)^2}\, \left[\Theta_{0,1}(2it)+\Theta_{1,1}(2it)\right] \cr
\times \int_{0}^{\infty} dP'\, \cos2\sqrt{2}\pi PP'\, q_{o}^{P'^2} \, \frac{\vartheta_{10}(2it)}{\eta^3(2it)} \,.
\end{multline}
\begin{multline}
\cK^{(\half)}_{\sR+\sR+}= \half\, 2^{4}\, \int \frac{ds}{128\pi^4}\, \frac{\vartheta_{10}(is)}{\eta^3(is)}\, \int_{0}^{\infty} dP\, \frac{\vartheta_{10}(2is)}{(\sinh \pi P)^2}\, q_{c}^{P^2}\, \frac{\vartheta_{10}(is)}{\eta^3(is)}\cr
= \half\, \int\frac{dt}{128\pi^4\, t^3}\, \frac{\vartheta_{01}(2it)}{\eta^3(2it)}\, \int_{0}^{\infty} \frac{dP}{(\sinh \pi P)^2}\, \left[\Theta_{0,1}(2it)-\Theta_{1,1}(2it)\right] \cr
\times \int_{0}^{\infty} dP'\, \cos2\sqrt{2}\pi PP'\, q_{o}^{P'^2} \, \frac{\vartheta_{01}(2it)}{\eta^3(2it)}  \,.
\end{multline}
where we used the identity
\begin{equation}
\vartheta_{01}(\tau)=\vartheta_{00}(4\tau)-\vartheta_{10}(4\tau) \,.
\end{equation} 
All contributions to the crosscap self overlap are singular as $P\rightarrow
 0$ and the
 total contribution to the Klein-bottle amplitude vanishes because of the identity 
\begin{align}
\vartheta_{00}(2is)\left(\frac{\vartheta_{00}^2(is)}{\eta^6(is)}-\frac{\vartheta_{01}^2(is)}{\eta^6(is)}\right) -\vartheta_{10}(2is)\frac{\vartheta_{10}^2(is)}{\eta^6(is)} &=0 \,.
\end{align}

The calculations  for the $J=0$ crosscap are similar to the ones done
above. We get: 
\begin{align}
\cK^{(0)}_{\sNS+\sNS+}&=2^{4}\, \int \frac{ds}{128\pi^4}\, \frac{\vartheta_{00}(is)}{\eta^3(is)}\,  \int_{0}^{\infty} dP\, \vartheta_{10}(2is)\, q_{c}^{P^2}\, \frac{\vartheta_{00}(is)}{\eta^3(is)}\cr
&= \int\frac{dt}{32\sqrt{2}\pi^4\, t^3}\, \frac{\vartheta_{00}(2it)}{\eta^3(2it)}\, \left[\Theta_{0,1}(2it)-\Theta_{1,1}(2it)\right]\, \frac{\vartheta_{00}(2it)}{\eta^3(2it)} \,.
\end{align}
\begin{align}
\cK^{(0)}_{\sNS+\sNS-}&= 2^4\,\int \frac{ds}{128\pi^4}\frac{\vartheta_{01}(is)}{\eta^3(is)}\,  \int_{0}^{\infty} dP\, \vartheta_{10}(2is)\, q_{c}^{P^2}\, \frac{\vartheta_{01}(is)}{\eta^3(is)}\cr
&= \int\frac{dt}{32\sqrt{2}\pi^4\, t^3}\, \frac{\vartheta_{10}(2it)}{\eta^3(2it)}\, \left[\Theta_{0,1}(2it)-\Theta_{1,1}(2it)\right]\, \frac{\vartheta_{10}(2it)}{\eta^3(2it)} \,.
\end{align}
\begin{align}
\cK^{(0)}_{\sR+\sR+}&=2^{4}\, \int \frac{ds}{128\pi^4}\, \frac{\vartheta_{10}(is)}{\eta^3(is)}\, \int_{0}^{\infty} dP\, \vartheta_{00}(2is)\, q_{c}^{P^2}\, \frac{\vartheta_{10}(is)}{\eta^3(is)}\cr
&= \int\frac{dt}{32\sqrt{2}\pi^4\, t^3}\, \frac{\vartheta_{01}(2it)}{\eta^3(2it)}\, \left[\Theta_{0,1}(2it)+\Theta_{1,1}(2it)\right]\, \frac{\vartheta_{01}(2it)}{\eta^3(2it)}\,.
\end{align}
Note that in the $t$-channel the density of states is a delta function as a result of which there is no integral over the momenta in the radial direction of the cigar. Thus, the self overlap is regular as $P\rightarrow 0$ and vanishes because of the identity (in the $s$-channel)
\begin{align}
\vartheta_{10}(2is)\left(\frac{\vartheta_{00}^2(is)}{\eta^6(is)}+\frac{\vartheta_{01}^2(is)}{\eta^6(is)}\right) -\vartheta_{00}(2is)\frac{\vartheta_{10}^2(is)}{\eta^6(is)} &=0 \,.
\end{align}

\end{appendix}

\end{document}